%% file: paper.tex
\documentclass[sigconf, screen, nonacm]{acmart}

\usepackage{graphicx}   
\usepackage{subcaption} 
\usepackage{soul}
\usepackage{footmisc}
\usepackage{flushend}
\usepackage{xfrac}
\usepackage{xspace}
\usepackage{nicefrac}
\usepackage{array}
\usepackage{diagbox}
\usepackage{enumitem}
\usepackage[normalem]{ulem}

\newcommand{\sref}[2]{\hyperref[#2]{#1 \ref{#2}}}

\newtheorem{thm}{Theorem}[section]

\newtheorem{dfn}[thm]{Definition}

\begin{document}








\title[Quantifying the Carbon Reduction of DAG Workloads: A Job Shop Scheduling Perspective]{Quantifying the Carbon Reduction of DAG Workloads:\\ A Job Shop Scheduling Perspective}




%



\author{
Roozbeh Bostandoost,
Adam Lechowicz,
Walid A. Hanafy,
Prashant Shenoy,
Mohammad Hajiesmaili\\[0.75ex]
\textit{University of Massachusetts Amherst}
}
\renewcommand{\shortauthors}{Bostandoost et al.}

\begin{abstract}

\input{0-abstract}

\end{abstract}

\maketitle

\section{Introduction}
\label{sec:introduction}
\input{1-introduction-v2}

\section{Problem Statement}
\label{sec:problem}

\input{2-problem}

\section{Evaluation}
\label{sec:eval}

\input{3-experiments}

\section{Discussion}
\label{sec:discussion}

\input{4-discussion-v4}



\clearpage
\bibliographystyle{ACM-Reference-Format}
\bibliography{paper}

\clearpage
\appendix
\input{appendix}

\end{document}

%% file: 0-abstract.tex
Carbon-aware schedulers aim to reduce the operational carbon footprint of data centers by running flexible workloads during periods of low carbon intensity. Most schedulers treat workloads as single monolithic tasks, ignoring that many jobs, like video encoding or offline inference, consist of smaller tasks with specific dependencies and resource needs; however, knowledge of this structure enables opportunities for greater carbon efficiency.

We quantify the maximum benefit of a dependency-aware approach for batch workloads. We model the problem as a flexible job-shop scheduling variant and use an offline solver to compute upper bounds on carbon and energy savings. Results show up to 25\% lower carbon emissions on average without increasing the optimal makespan (total job completion time) compared to a makespan-only baseline. Although in heterogeneous server setup, these schedules may use more energy than energy-optimal ones. Our results also show that allowing twice the optimal makespan nearly doubles the carbon savings, underscoring the tension between carbon, energy, and makespan. We also highlight key factors such as job structure and server count influence the achievable carbon reductions.



%% file: 1-introduction-v2.tex
In recent years, rising compute demand and data center buildings have raised concerns about electricity use and the resulting \textit{operational} carbon emissions of ICT~\cite{GoldmanSachs:25, Shehabi2024:USDCReport}.
To address these concerns, 
data center operators improved energy efficiency for more than a decade~\cite{Barroso_warehouse}, but these improvements are approaching fundamental limits with the advent of hyperscale data centers~\cite{Bashir:23}. Thus, future growth in demand~\cite{Shehabi2024:USDCReport} is expected to increase energy usage significantly~\cite{Shehabi2024:USDCReport} -- since the favored choices for new power generation to support data centers continue to be fossil-based resources such as natural gas turbines~\cite{NYT:25GasTurbines, skidmore2025:Stargate}, this projected growth is expected to correlate with increases in carbon emissions.

To address concerns about carbon footprint, operators turned to \textit{supply-side} optimizations, e.g., by purchasing renewable power or co-locating data centers with wind and solar farms~\cite{Acun:23, GoogleRenewableEnergy:18, Calma:24}. Researchers have also proposed \textit{demand-side} techniques 
to reduce the operational carbon of batch workloads using temporal or spatial shifting~\cite{wait-awhile, sukprasert:2023:quantifying, hanafy:2023:carbonscaler, Zheng:20, lin2023adapting, Gsteiger:24}.  
These works rely on visibility into the grid's \textit{carbon intensity} (i.e., amount of carbon emitted per unit of energy used) and flexible applications that can modulate their resource consumption in response to variations on the grid.
Existing work ranges from theoretical models of single jobs with deadlines~\cite{Lechowicz:23} to experimental systems that provision resources for many jobs in response to carbon intensity~\cite{radovanovic2022carbon, souza:2023:ecovisor, Acun:23, Chadha:23, Gsteiger:24, perotin2023risk}.

\begin{figure}[h]
    \centering
    \includegraphics[width=0.8\linewidth]{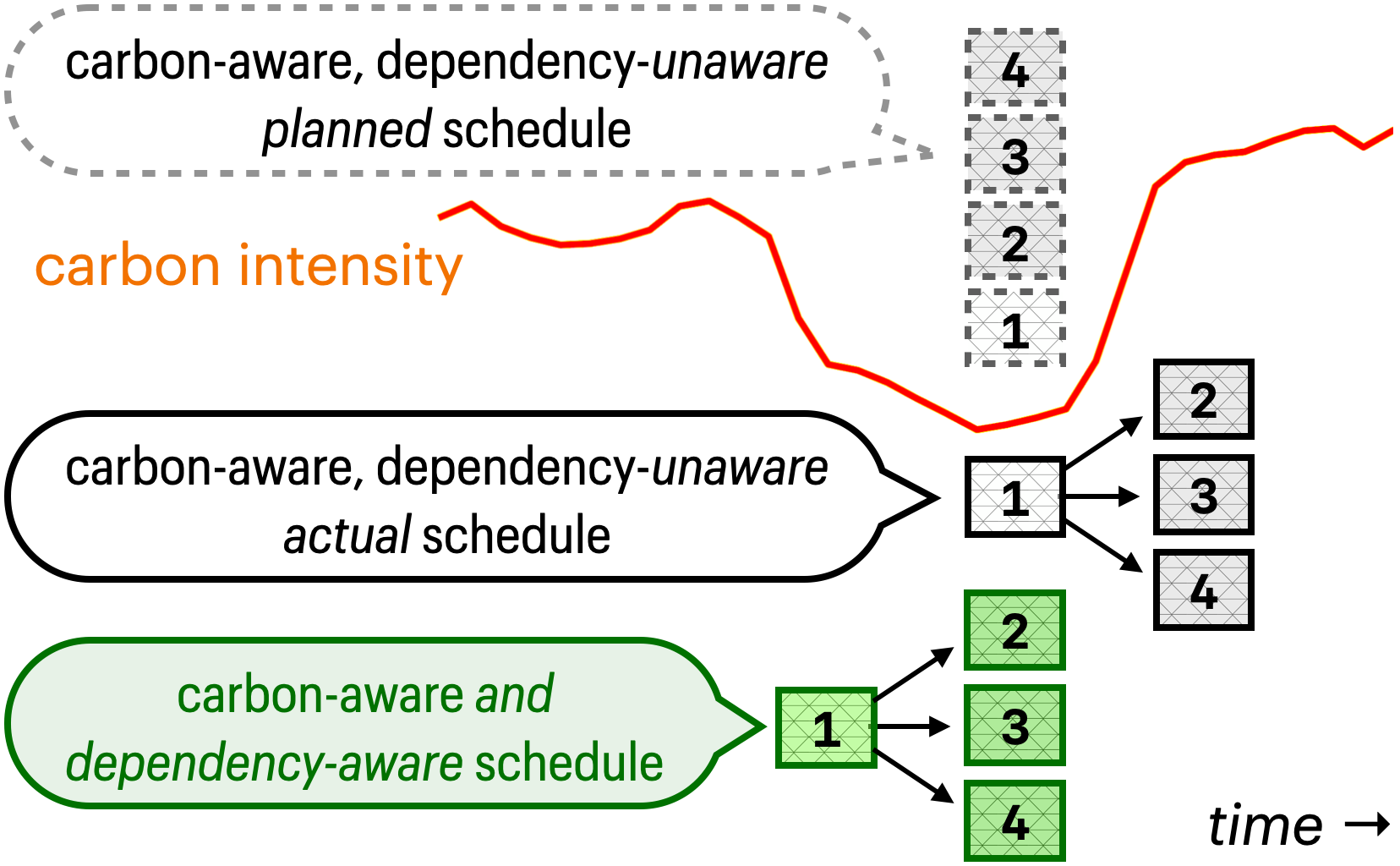} \vspace{-1em}
\caption{A dependency-unaware scheduler attempts to run all tasks in the low-carbon period, but precedence constraints force most tasks to run during a high-carbon period. A dependency-aware scheduler makes a better trade-off for a lower total carbon footprint. }
    \label{fig:motivation-dag} \vspace{-1em}
\end{figure}

Existing carbon-aware schedulers typically treat each workload as a single, monolithic task. This model overlooks a key reality: Many workloads decompose into dependent \textit{tasks} with different resource and energy profiles, often across heterogeneous machines (e.g., CPU vs.\ GPU, hardware generations). 
Examples of these workloads include video encoding/transcoding, serverless pipelines~\cite{Kounev2023:Serverless}, storage management~\cite{Rao:25}, and offline inference~\cite{Alibaba:19, wei2022status}. Recognizing this structure opens new opportunities for carbon efficiency. As \autoref{fig:motivation-dag} shows, a dependency-unaware scheduler may force tasks into high-carbon periods, negating potential savings.

While recent workflow systems (e.g., Caribou~\cite{Gsteiger:24}) begin to model multi-stage dependencies as directed acyclic graphs (DAGs) in the serverless setting, data center operators need to know the maximum possible (carbon or energy) savings before investing in complex online solutions. This paper quantifies the upper bound of the (carbon or energy) savings by assuming perfect offline knowledge of job arrivals, task dependencies (in the format of DAGs), and their energy profiles. Our analysis focuses on batch workloads as they represent a significant portion of data center computation.

To quantify the potential savings of temporal shifting for DAG jobs, we model the problem as a variant of the flexible job-shop scheduling problem (FJSP)~\cite{Dauzere:24}. Our formulation optimizes for energy or operational carbon under a makespan constraint and compares against the standard makespan objective. 
We generate batch-job DAGs inspired by Alibaba traces~\cite{Alibaba:18}, pair them with Electricity Maps carbon traces~\cite{electricity-map}, and use a constraint solver~\cite{ortools} to explore feasible schedules. Our contributions are as follows:

\begin{enumerate}[leftmargin=*]
    \item Formulate a carbon-aware flexible job shop problem with inter-task dependencies and compute upper bounds on achievable carbon savings. Our results highlight that carbon-aware scheduling can lower carbon emissions by up to 25\% on average without increasing the optimal makespan.
    \item Quantify trade-offs between carbon, energy, and makespan. In heterogeneous server setup, we show that carbon-optimal schedules can use up to 7\% more energy than energy-optimal schedules with the same makespan, and that doubling the makespan flexibility nearly doubles carbon savings.
    \item Identify parameters that drive carbon savings: more tasks per job raises server utilization and cuts savings by up to 35\%, while adding servers lowers server utilization and yields up to $30\times$ higher savings.
\end{enumerate}

%% file: 2-problem.tex
In this section, we present the flexible job shop scheduling problem (FJSP) in the computing context of tasks with dependencies before formalizing the main objectives (makespan, energy, carbon) that we consider in the rest of the paper.  We give brief background on the history of FJSP, including some recent works that have expanded the study of the problem beyond the singular objective of makespan.

We consider a set of jobs denoted by $\mathcal{J}$, where each job $j \in \mathcal{J}$ consists of $n_j$ tasks $\{ t_{j, 1}, \cdots,  t_{j, n_j} \}$.  Each job also has a dependency graph $G$ (as a DAG) with directed edges $\mathcal{E}$ between tasks; for instance, if the edge $(k-1, k)$ exists in $E$, this indicates that task $t_{j, k}$ cannot run until task $t_{j, k-1}$ completes.

Given a finite set of machines $\mathcal{M}$ that may be heterogeneous, any task $t$ can run on a subset of machines denoted by $\mathcal{M}_t$, with instantaneous energy usage $E_{t, m}$ and processing time $p_{t, m}$ on machine $m \in \mathcal{M}_t$.
Finally, to model realistic scenarios in a data center, each job has an arrival time $a_j \geq 0$.  If $a_j > 0$, job $j$'s first task cannot begin until after time $a_j$.



A feasible schedule $\mathcal{S}$ assigns each task to a machine and a start time, respecting task dependencies and machine capacity. For the $i$th task in job $j$, we denote the start time as $s_{j,i}$ and $x_{j,i,m} \in \{0,1\}$ equals 1 if task $t_{j,i}$ runs on machine $m$. (see \autoref{sec:problem-formulation} for the problem formulation). We evaluate three objectives:
\begin{dfn}[Makespan]\label{dfn:mkspan}
    The \textbf{makespan} of a schedule is defined as $\max_{j \in \mathcal{J}} C_{j, n_j}$, where $C_{j, n_j}$ is the completion time of job $j$ (i.e., the time at which task $t_{j, n_j}$ completes).
\end{dfn}
\begin{dfn}[Energy Usage]\label{dfn:energy}
    The \textbf{energy usage} of a schedule is given by $\sum_{j \in J} \sum_{i=1}^{n_j} \int_{s_{j, i}}^{C_{j, i}} E_{t_{j, i}, m} \, d \tau$, where $E_{t_{j, i}, m}$ is the energy usage of task $i$ belonging to job $j$ on the assigned machine $m$.
\end{dfn} 
\begin{dfn}[Carbon Emissions]\label{dfn:carbon}
    The \textbf{ carbon emissions} of a schedule are quantified by $\sum_{j \in J} \sum_{i=1}^{n_j} \int_{s_{j, i}}^{C_{j, i}} E_{t_{j, i}, m} \cdot I(\tau) \, d \tau$, where $I(\tau)$ denotes the grid carbon intensity at time $\tau$.
\end{dfn}
In most of the literature on JSP and related problems, the studied metric is \textit{makespan}:  the total time required to complete all jobs according to a given schedule.  This is distinct from e.g., average job completion time, which measures the time elapsed between a single job's submission and its completion.  
Since we are motivated by background batch jobs such as video encoding that are often centrally orchestrated by or ``belong to'' the data center operator, we focus on makespan in the interest of ensuring that all daily background jobs are completed as a group in a timely fashion.
However, we note that other performance metrics such as job completion time may be more appropriate in e.g., multi-tenant scenarios. 



\begin{figure}[h]
    \centering 
    \includegraphics[width=0.95\linewidth]{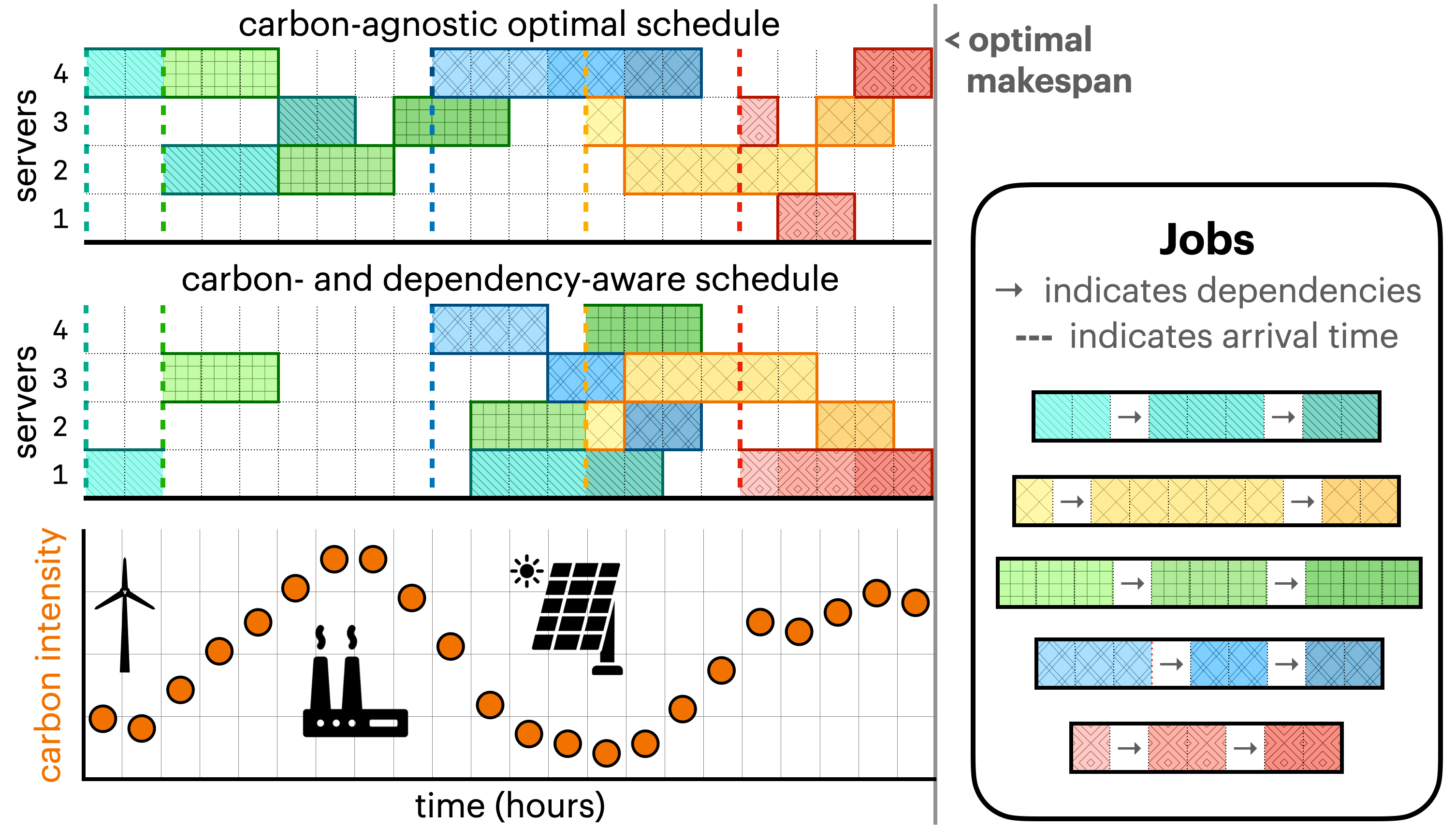} \vspace{-1em}
    \caption{An example FJSP with homogeneous machines. The top schedule is optimized for the fastest completion time (makespan). The middle schedule is optimized for the lowest carbon footprint, shifting tasks to align with low-carbon periods (bottom) while respecting the makespan constraint.}
    \label{fig:motivation} 
\end{figure}

To solve the FJSP under these objectives, we first solve the problem for the classic case of unconstrained makespan minimization (i.e., not minimizing energy or carbon), obtaining the optimal feasible makespan $\texttt{OPT}$.  From this ``baseline'', we then solve the problem with either energy or carbon as an objective, formalizing the makespan as a constraint (e.g., for a given \textit{stretch factor } $S \geq 1$, the second-level schedule's makespan must be at most $S \times \texttt{OPT}$).  We give a toy example of this bi-level optimization in \autoref{fig:motivation}, and detail our solver implementation in \autoref{sec:eval-setup}.

\smallskip
\noindent \textbf{Example Job: Offline Inference. \ }
To motivate how FJSP models important problems in the data center, we consider the example of offline inference, where a collection of new data is processed by a trained ML model to generate new predictions at regular intervals (e.g., daily)~\cite{MLRunBatchInference:25}. 
%
%
In an offline inference job, there are three distinct tasks to be executed in sequence, each using different resources and with different energy footprints.  First, one task loads the new inference data to be processed in batch using e.g., Hadoop or a Spark DataFrame~\cite{Spark:16}.  Then, the next task initializes and uses the ML model to predict outputs for each row in the batched data.  Finally, the last task stores and visualizes the resulting predictions for future use.  Note that each of these tasks involves different operations, and may even require different hardware (e.g., if the ML model is designed for deployment on a GPU cluster). 
Depending on the complexity of the model and the size of the data, these jobs can require hours or even days of runtime~\cite{SrinivasanParker:21}.  For instance, using the smallest No Language Left Behind model for a language translation task can require at least 480 hours of runtime, with 440 of those counting as GPU-hours~\cite[Table 15]{Koishekenov:23}. Other examples, include batch processing and HPC applications such as DNA sequencing~\cite{DNAp}, drug discovery~\cite{drugdiscovery}, web crawling~\cite{crawling}.



\smallskip
\noindent \textbf{Background. \ }
The job shop scheduling problem (JSP) models each job as a sequence of tasks on specific machines, aiming to minimize the overall makespan. JSP is known to be NP-hard~\cite{Garey:76}, so studies have focused on simple heuristics (e.g., list scheduling~\cite{Graham:66}) and approximation algorithms (see \citet{Xiong:22} for a survey). The flexible job shop problem (FJSP)~\cite{Brucker:90} generalizes JSP by allowing each operation to be assigned to any machine in a candidate set.
Although the JSP and FJSP have roots in operations research and manufacturing contexts, they are also relevant to computer systems scheduling,
where jobs arrive online and must be dispatched in real-time (e.g., using heuristics like earliest-task-first~\cite{coleman:25:pisaadversarialapproachcomparing}). 

In recent years, some studies have considered competing priorities, such as time-varying costs of production and/or energy usage~\cite{Terbrack:25, Para:22}. For instance, \citet{Park:22} considers the FJSP under time-of-use electricity pricing, aiming to minimize both the makespan and total energy cost.  They solve the problem optimally and show it is possible to significantly cut energy costs without compromising productivity. 
However, most of these works ignore carbon emissions and are grounded in manufacturing or industrial contexts.
Thus, in the rest of the paper, we explore the solution space of schedules for the FJSP in a setting tailored for computing workloads, using tools developed in operations research.

%% file: 3-experiments.tex
In this section, we describe our experimental setting and solution approach before presenting our main results.  Our code and data is available at \href{https://github.com/rbostandoust/HotCarbon25}{GitHub}. 

\subsection{Experimental Setup} \label{sec:eval-setup}
Using the formulation from \autoref{sec:problem}, we construct flexible job shop scheduling instances and solve them as follows.

\noindent \textbf{Server setup. \ }
We schedule each FJSP instance on $M$ servers, with $M=5$ unless noted. 
We consider two server setups that model homogeneity and heterogeneity, respectively.
In the \emph{homogeneous} setup, all the servers draw $1$kW and run at the same speed. 
In the \emph{heterogeneous} setup, we use 5 server classes with power draws of $0.25$, $0.5$, $1$, $1.5$, and $2$kW. The $1$kW server is the baseline with the same speed as in the homogeneous case. The servers run at $\{\nicefrac{1}{3}, \nicefrac{1}{2}, 1, \nicefrac{4}{3}, 2\}\times$ this baseline speed.

\begin{figure*}[t]
\vspace{-1em}
\begin{minipage}{0.29\linewidth}
    \begin{subfigure}[b]{0.49\textwidth}
        \centering
        \includegraphics[width=1\textwidth]{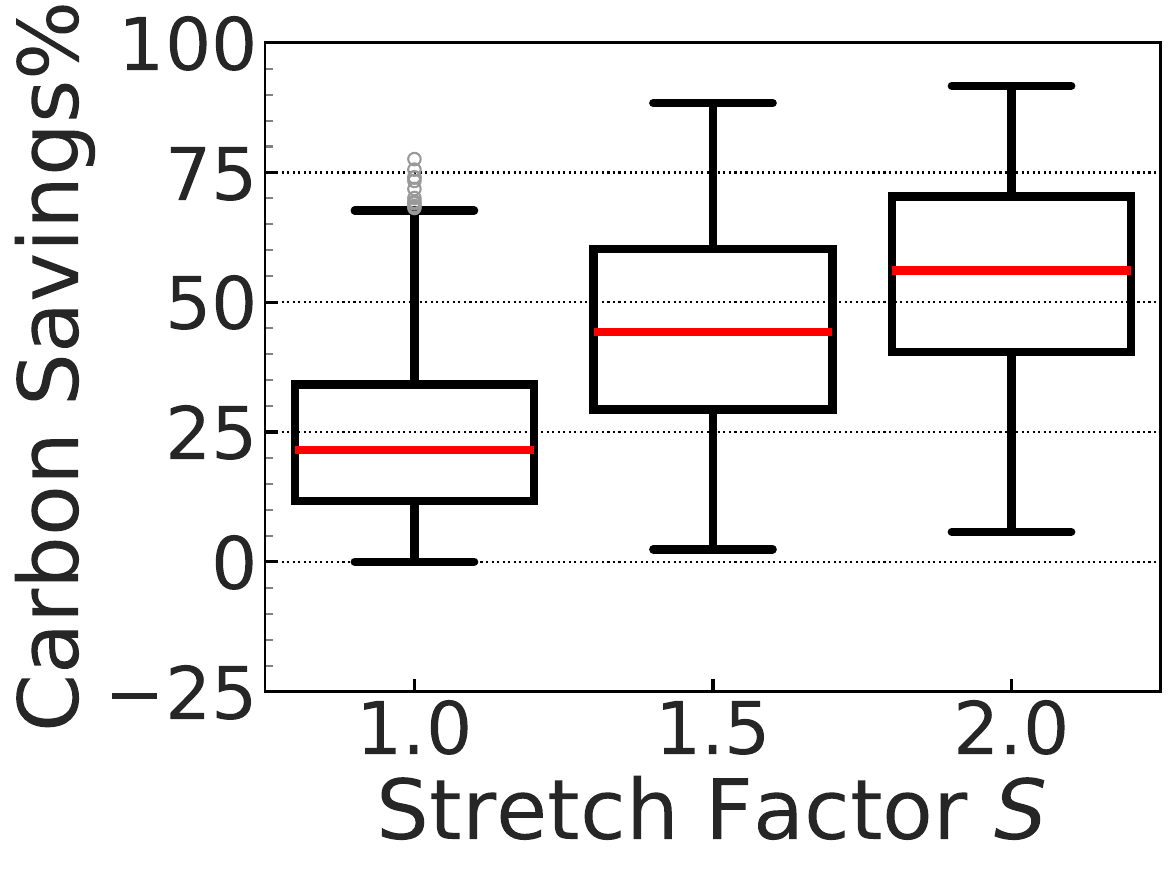}\vspace{-0.75em}
        \captionsetup{justification=centering}
        \caption{Homogeneous}
    \label{fig:homogen-main}
    \end{subfigure}
    \begin{subfigure}[b]{0.49\textwidth}
        \centering
        \includegraphics[width=1\textwidth]{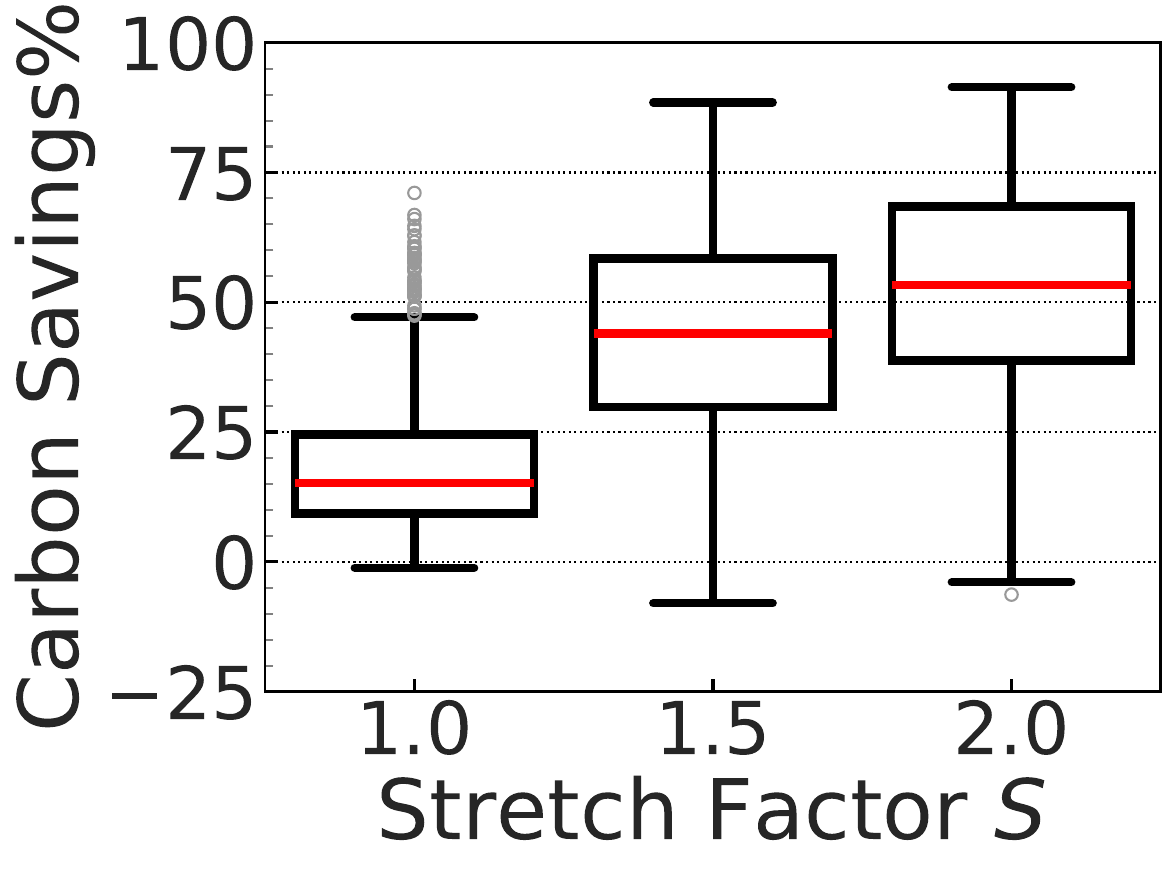}\vspace{-0.75em}
        \captionsetup{justification=centering}
        \caption{Heterogeneous}
    \label{fig:heterogen-main}
    \end{subfigure}\vspace{-1em}
    \caption{Carbon savings in the \texttt{AU-SA} grid at different stretch factors $S$. (a) Homogeneous servers. (b) Heterogeneous servers.}
    \label{fig:main-result}
\end{minipage} \hfill
\begin{minipage}{0.37\linewidth}
\centering
\begin{subfigure}[b]{0.49\textwidth}
    \centering
    \includegraphics[width=\textwidth]{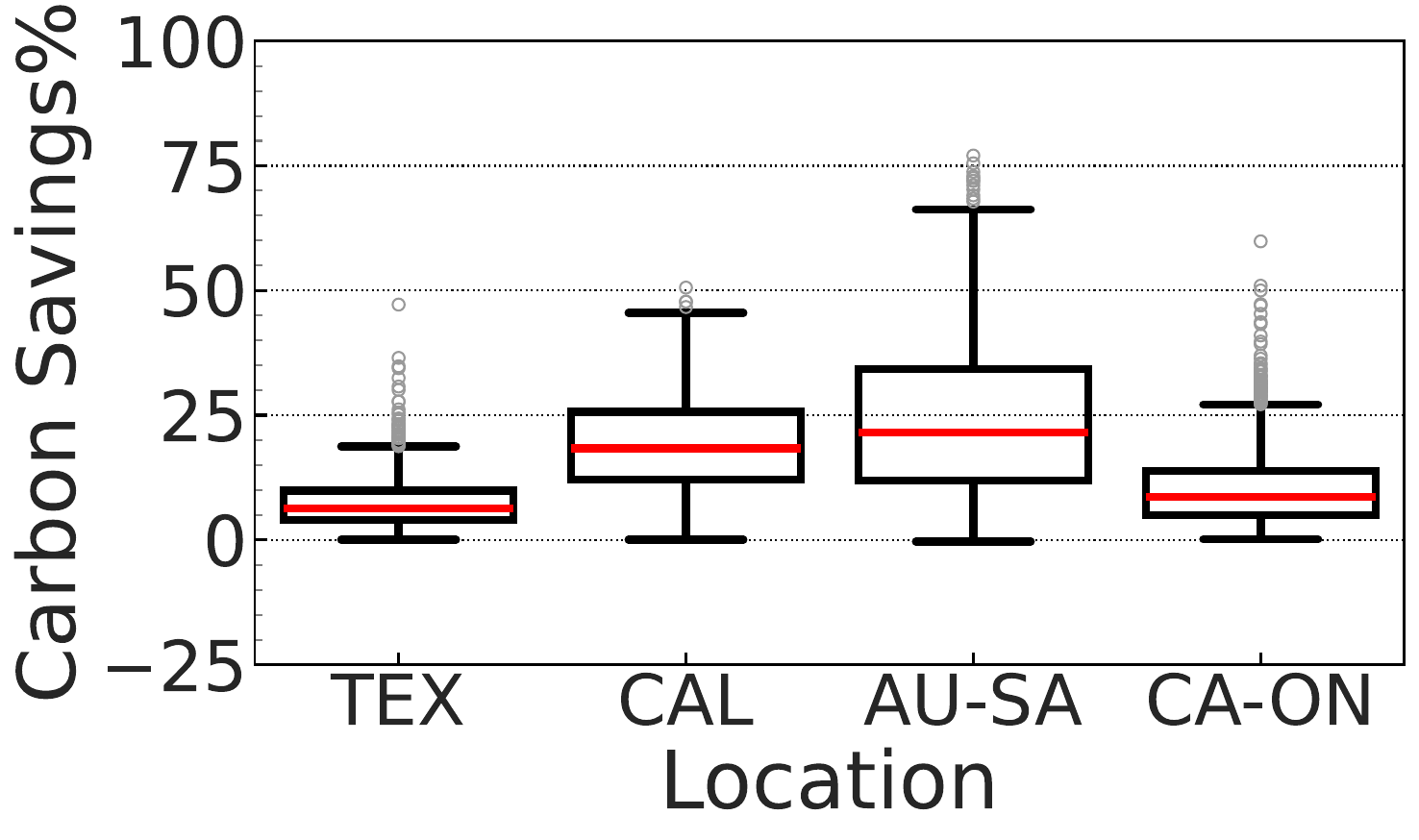}\vspace{-0.75em}
    \captionsetup{justification=centering}
    \caption{Homogeneous}
\end{subfigure}
\begin{subfigure}[b]{0.49\textwidth}
    \centering
    \includegraphics[width=\textwidth]{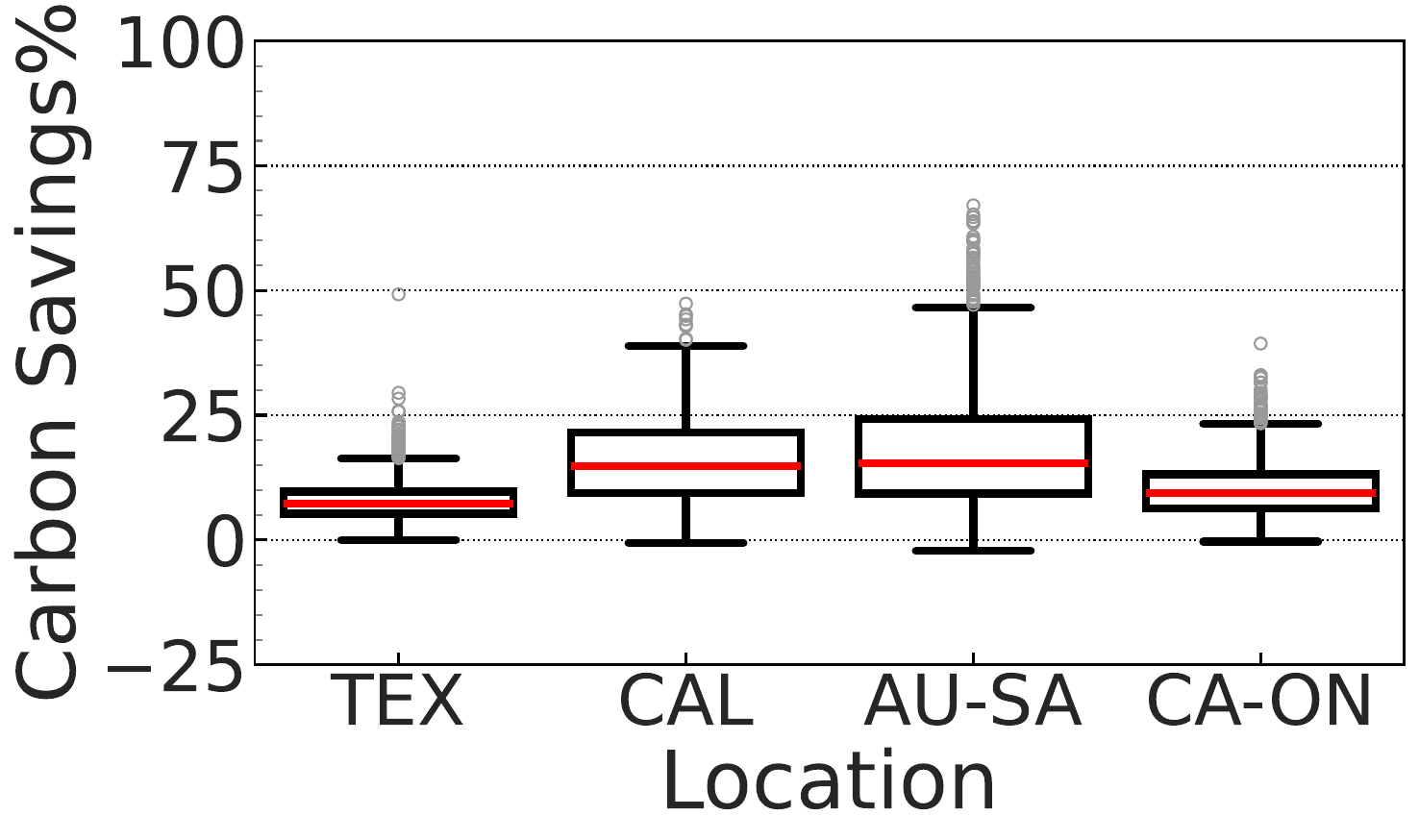}\vspace{-0.75em}
    \captionsetup{justification=centering}
    \caption{Heterogeneous}
\end{subfigure}\vspace{-1em}
\caption{Carbon savings achieved for $S=1$ in 4 different grid regions. (a) shows results for homogeneous case, while (b) shows results for the heterogeneous case. }
\label{fig:effect-of-regions}
\end{minipage} \hfill
\begin{minipage}{0.29\linewidth}
    \begin{subfigure}[b]{0.49\textwidth}
        \centering
        \includegraphics[width=\textwidth]{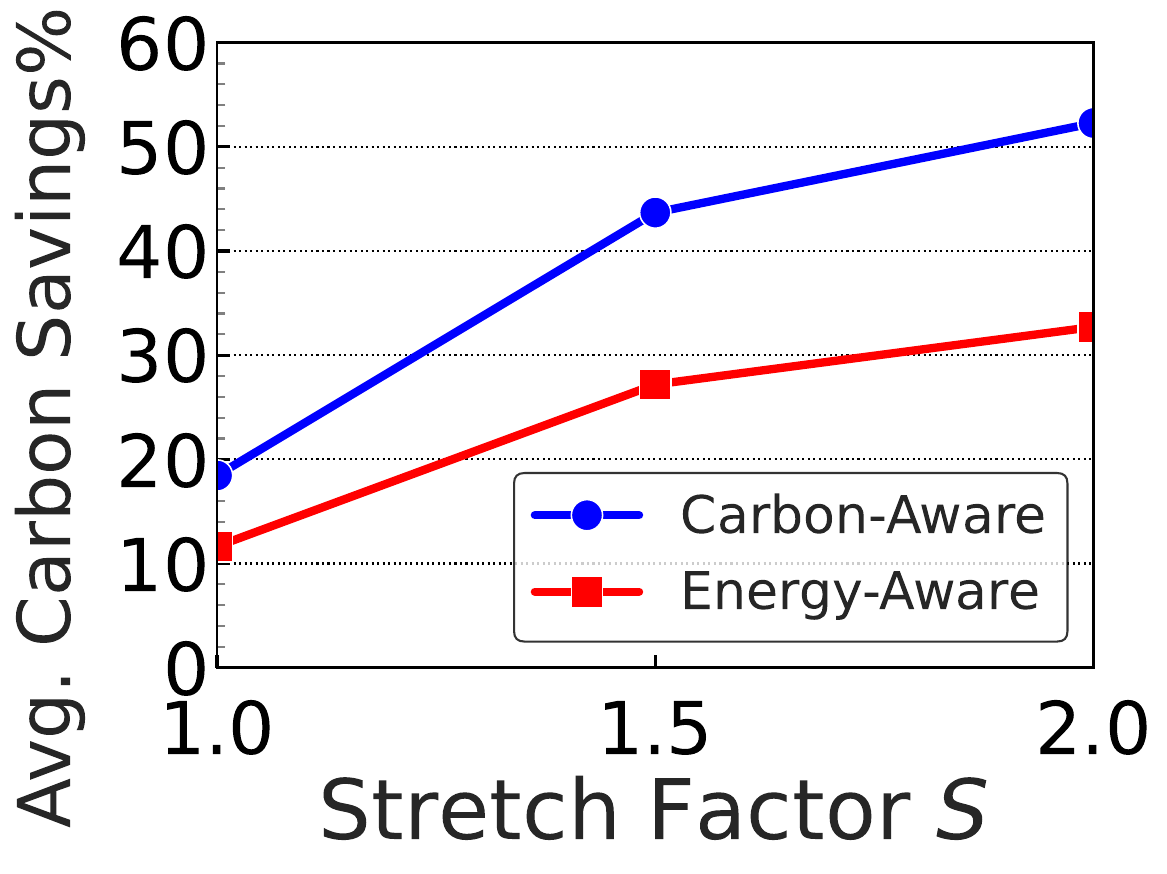}\vspace{-0.75em}
        \caption{Carbon savings}
        \label{fig:carbon-vs}
    \end{subfigure}\hfill
    \begin{subfigure}[b]{0.49\textwidth}
        \centering
        \includegraphics[width=\textwidth]{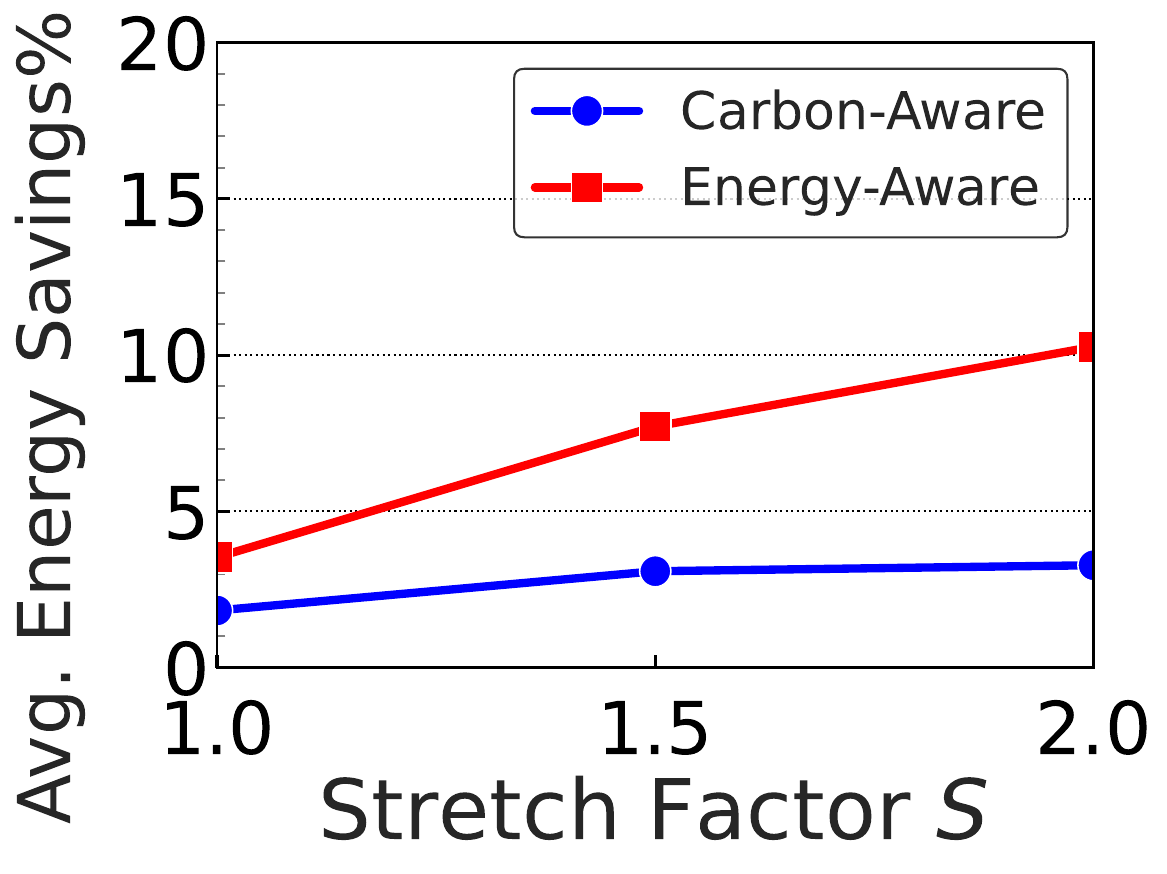}\vspace{-0.75em}
        \caption{Energy savings}
        \label{fig:energy-vs}
    \end{subfigure} \vspace{-1em}
    \caption{Comparing the carbon savings and energy savings achieved in \texttt{AU-SA} grid by solvers that optimize for carbon and energy, respectively.}
    \label{fig:carbon-vs-energy} 
\end{minipage}\vspace{-1em}
\end{figure*}

\noindent \textbf{Carbon intensity data.} 
We use hourly carbon intensity data from Electricity Maps~\cite{electricity-map}, mainly from the South Australia region (\texttt{AU-SA}) in 2024. This region shows high daily variation and strong penetration of renewables, making it a good case for carbon-aware scheduling~\cite{sukprasert:2023:quantifying}. Results for other regions appear in \autoref{fig:effect-of-regions}. 
Each instance starts at a random point in the trace ($t=0$). Each of the $n$ jobs has a uniform random arrival time in the next 24 hours. 
We discretize time into 15-minute epochs, rounding each task’s processing time up to reduce solver complexity.

\begin{figure}[h]
    \centering\vspace{-1em}
    \includegraphics[width=0.98\linewidth]{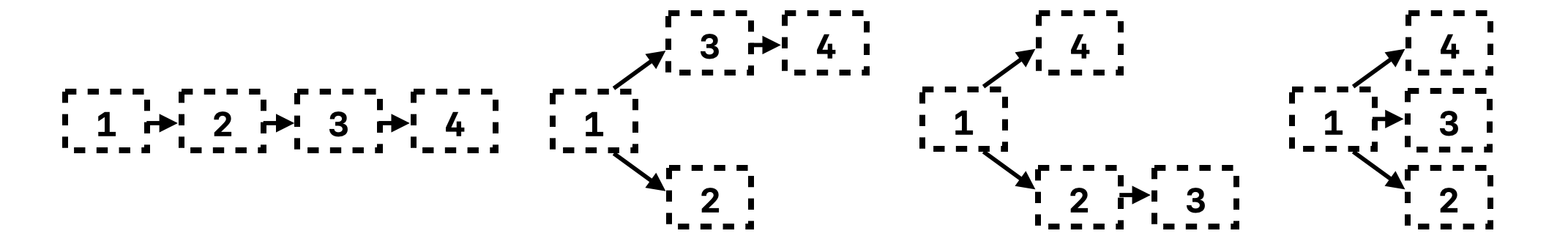} \vspace{-1em}
    \caption{
    For the case of $k=4$ tasks, the dependency structures that we sample from in our experiments.
    }
    \label{fig:dependencies} \vspace{-1em}
\end{figure}

\noindent \textbf{Jobs and task dependency structures. \ }
Each instance has $n$ jobs, each with $k$ tasks. Unless noted, we set $n=10$ and $k=4$. Jobs follow simple dependency graphs: a chain, two branches from a root, or one root feeding all other tasks (\autoref{fig:dependencies}). 
In the homogeneous setup, all servers run tasks at the same speed with the same energy use. We sample task durations from an exponential distribution with mean $\lambda=7$ epochs, which matches the average length of tasks that are longer than one minute in the Alibaba trace~\cite{Alibaba:18}. This models the long batch jobs we seek to address.


In the heterogeneous setup, task times scale with server class. For example, a task that takes $10$ epochs on the 1kW baseline server takes $\{30, 20, 10, 7.5, 5\}$ epochs on the five server types, in line with their relative speeds.

\begin{figure}[h]
\vspace{-1em}
\begin{minipage}{0.4\linewidth}
    \includegraphics[width=\linewidth]{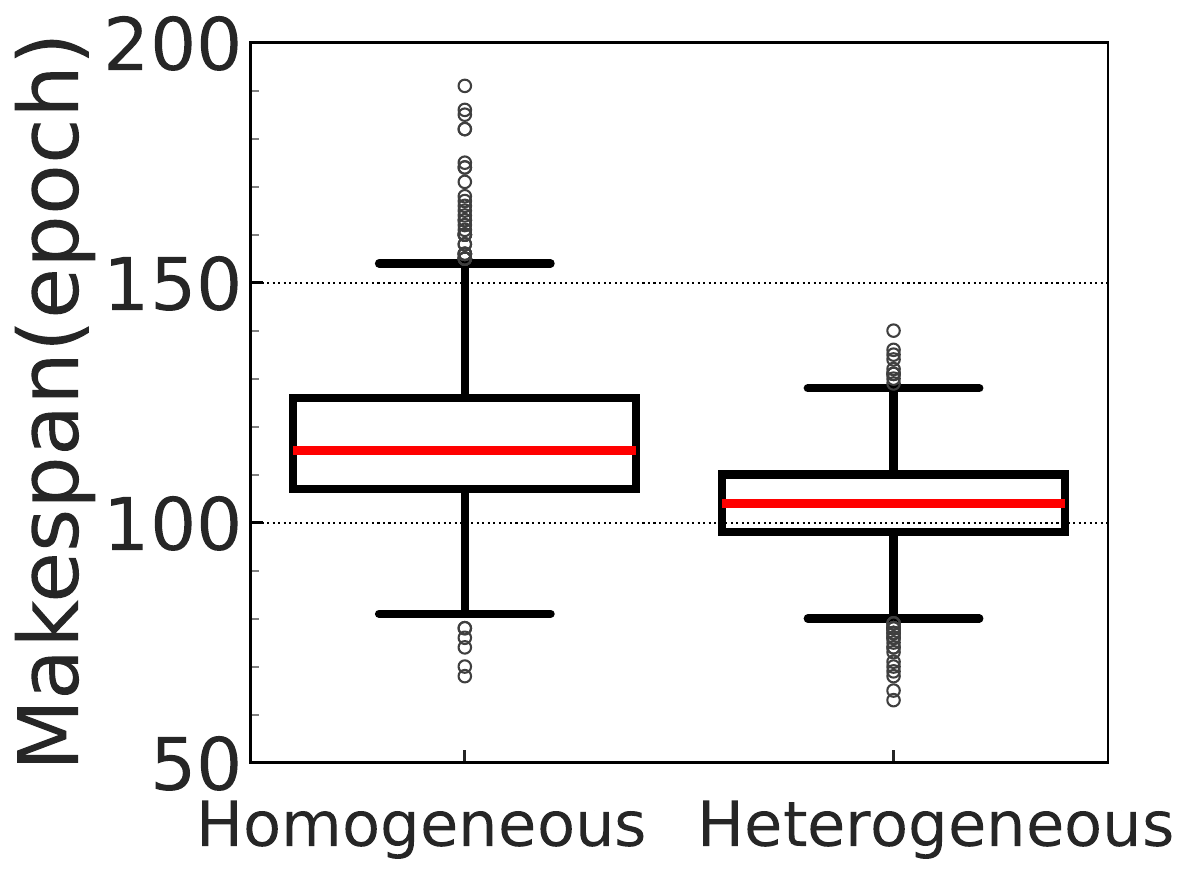}
\end{minipage}\hfill
\begin{minipage}{0.55\linewidth}\vspace{-1em}
    \caption{Optimal makespan (in epochs, $1$ epoch = 15 min.) for homogeneous and heterogeneous server settings in our experiments.  $n = 10$ jobs, $M = 5$ servers, and each job has $k = 4$ operations. }\label{fig:makespan}
\end{minipage}\vspace{-1em}
\end{figure}

\noindent \textbf{Solver implementation. \ }
We model FJSP with constraint programming in Google OR-Tools~\cite{ortools, cpsatlp}. 
For each instance, we first solve for the optimal makespan using the standard FJSP model, ignoring carbon or energy optimization. \autoref{fig:makespan} shows the distribution of this optimal makespan in the homogeneous and heterogeneous setups.
We then re-solve each instance with objectives of minimizing energy or operational carbon consumption. In these runs, we add a makespan constraint with a stretch factor $S \geq 1$. If the optimal makespan is 10 epochs, $S=2$ allows any schedule that finishes within 20 epochs. Larger $S$ values give the solver more room to shift tasks into low-carbon periods, but also make the search space and problem complexity grow. To keep runtimes manageable, we set solver timeouts of $1$, $3$, and $5$ minutes for $S=1$, $1.5$, and $2$, respectively.


\subsection{Experimental Results }

\noindent \textbf{Carbon emissions vs. makespan trade-off. \ }
We study how much carbon-aware scheduling can reduce emissions compared to a baseline that only minimizes makespan (carbon-agnostic). Carbon savings are reported relative to this baseline schedule.
There is a trade-off between carbon savings and job completion time: more slack gives the solver more freedom to move tasks into low-carbon periods~\cite{sukprasert:2023:quantifying}. We evaluate 1000 instances from the \texttt{AU-SA} grid region. \autoref{fig:main-result} shows the results for homogeneous and heterogeneous setups.

With no slack ($S=1$), the carbon-aware scheduler cuts emissions by an average of 25\% in the homogeneous case and 18\% in the heterogeneous case. Allowing more slack boosts these gains: At $S=2$, average savings rise to 54\% and 52\% respectively.
In the heterogeneous setup (\autoref{fig:heterogen-main}), the solver occasionally finds worse solutions than the carbon-agnostic one, resulting in negative savings. This happens because a larger $S$ increases the complexity of the problem, and the added complexity makes it harder for the solver to find an optimal solution within the time limit.

\autoref{fig:main-result} also shows that the heterogeneous setup achieves lower carbon savings than the homogeneous one. The scheduler has less flexibility to shift tasks due to two factors: a shorter overall makespan (\autoref{fig:makespan}) and higher average server utilization. For example, when $S=1$, the average utilization is 55.02\% in the heterogeneous case versus 47.15\% in the homogeneous case.




\smallskip
\noindent \textbf{Effect of carbon trace. \ } 
Carbon savings depend heavily on the grid region~\cite{sukprasert:2023:quantifying}. \autoref{fig:effect-of-regions} shows achieved carbon savings for $S=1$ using 2024 data from California (\texttt{CAL}), Ontario (\texttt{CA-ON}), and Texas (\texttt{TEX}), in addition to South Australia (\texttt{AU-SA}).
We find large differences. Texas shows smaller savings because its carbon intensity varies less during the day and is higher on average than in California or South Australia. Ontario shows the opposite: while variation is high, its average carbon intensity is already very low (about 90\% low-carbon), so the room for improvement is limited. These contrasts highlight how both the variability and the baseline level of carbon intensity shape the benefits of carbon-aware scheduling.



\smallskip

\noindent \textbf{Effect of number of servers. \ }  
We evaluate the effect of server count $M=\{2,5,10\}$ in the homogeneous setting with $n=10$ jobs and $k=4$ operations per job. \autoref{tab:num-servers} reports results for $S=1$ in the \texttt{AU-SA} grid, averaged over 1000 instances.  

More servers give the scheduler more room to place tasks in low-carbon periods. Moving from 2 to 10 servers increases savings by up to $30\times$. A larger server pool also lowers server average utilization, since the same workload spreads across more machines. This increases parallelism and shortens the makespan. But with only 10 jobs, the gain from 5 to 10 servers is negligible, and the makespan hardly improves.

\smallskip

\noindent \textbf{Effect of number of tasks. \ }  
We vary the number of tasks per job with $k=\{3,4,5\}$ while keeping other parameters fixed. \autoref{tab:num-ops} reports results for $S=1$ in the \texttt{AU-SA} grid, averaged over 1000 instances.  
More tasks reduce the achievable carbon savings. With three tasks per job, average savings are 30.4\%, but with five tasks, the savings drop to 19.7\%. The reason is higher server utilization: as $k$ increases from 3 to 5, average server utilization rises from 36.4\% to 56.6\%. Higher server utilization leaves the solver less flexibility to shift workloads into low-carbon periods.

\smallskip
\noindent \textbf{Optimizing for energy usage vs. carbon emissions. \ }
Operators face a fundamental choice between energy efficiency and carbon efficiency as the primary objective when scheduling batch workloads. 
Energy efficiency pushes workloads onto the most power-efficient servers, maximizing throughput per watt. 
Carbon efficiency shifts workloads into time windows when the grid is cleanest. 
These goals can conflict: the most efficient server may be active during a high-carbon period, while a less efficient server running later could yield lower emissions overall.
To quantify this trade-off, 
we configure our makespan-constrained solver to prioritize energy reduction (\sref{Def.}{dfn:energy}). Also, to implement tie-breaking rules, we assign a small weight to carbon emissions, so when two schedules use the same energy, the solver prefers the one with lower carbon emissions.


\begin{table}[t]
\begin{minipage}{1\linewidth}
\setlength{\tabcolsep}{3pt}
    \footnotesize
    \caption{Average carbon savings, makespan, and utilization for varying instance parameters in \texttt{AU-SA} grid, with $S=1$ and homogeneous servers.} \vspace{-1em}
    \label{tab:effect-of-servers-ops}
    \begin{subtable}[b]{0.48\textwidth}
        \centering
        \caption{No. of servers $M$}
        \label{tab:num-servers}
        \begin{tabular}{|c|c|c|c|}
        \hline
        $M$ & 2                          & 5                          & 10                         \\ \hline
        \begin{tabular}[c]{@{}l@{}}Carbon\\ Savings\end{tabular}    & 1.13\% & 24.64\% & 33.98\% \\ \hline
        \begin{tabular}[c]{@{}l@{}}Makespan\\ (epochs)\end{tabular} & 153.74   & 117.73   & 117.63   \\ \hline
        Utilization & 89.39\% & 47.15\% & 23.6\% \\ \hline
        \end{tabular}
    \end{subtable} \hfill
    \begin{subtable}[b]{0.48\textwidth}
        \centering
        \caption{No. of tasks per job $k$}
        \label{tab:num-ops}
        \begin{tabular}{|c|c|c|c|}
        \hline
        $k$ & 3                          & 4                          & 5                         \\ \hline
        \begin{tabular}[c]{@{}l@{}}Carbon\\ Savings\end{tabular}    & 30.43\% & 24.64\% & 19.69\% \\ \hline
        \begin{tabular}[c]{@{}l@{}}Makespan\\ (epochs)\end{tabular} & 110.42   & 117.73   & 121.77   \\ \hline
        Utilization & 36.39\% & 47.15\% & 56.61\% \\ \hline
        \end{tabular}
    \end{subtable}
\end{minipage} \hfill
\vspace{-.5cm}
\end{table}

We evaluate heterogeneous servers and average results over 1000 instances in the \texttt{AU-SA} grid region. We first use a solver as the baseline to compute the optimal makespan, then solve a constrained FJSP with carbon or energy as the primary objective. Savings are reported relative to the baseline’s energy or carbon consumption. \autoref{fig:carbon-vs-energy} shows the results. The energy-focused solver reduces carbon emission by $\sim30\%$ when $S=2$, but the carbon-focused solver achieves $\sim50\%$ savings. For energy efficiency, the pattern reverses: the energy-focused solver achieves about $10\%$ savings at $S=2$, while the carbon-focused solver achieves only about $3\%$.


These results highlight a \textit{tension} between energy usage and carbon emissions. These two objectives align when saving energy occurs during a low-carbon period. However, they conflict if an energy-saving strategy extends a task into a high-carbon period, which can increase overall carbon emissions.

%% file: 4-discussion-v4.tex
We quantified \textit{upper bounds} on the achievable carbon reduction for workloads with inter-task dependencies in offline settings, illustrating a clear trade-off between carbon, energy, and makespan. These findings may inform the design of online heuristics to solve the problem in practice, and can give intuition about the preferrable ``trade-off point'' for relevant applications.


\noindent\textbf{Carbon savings without makespan loss.} Compared to a carbon-agnostic, makespan-minimizing schedule, a carbon-aware schedule can reorder tasks to exploit low-carbon periods. This works because servers in the makespan-optimal schedule may sit idle in low-carbon periods: using those idle gaps, a scheduler can increase server utilization in low-carbon periods, cutting emissions without extending the makespan (\autoref{fig:motivation}). 

\noindent\textbf{Carbon vs. makespan.} Scheduling purely for carbon can increase makespan, so both metrics must be considered together. Our results show diminishing returns: allowing makespan of $1.5\times$ the optimal achieves most of the carbon savings, while further relaxation yields little additional savings (\autoref{fig:main-result}).  

\noindent\textbf{Carbon vs. energy.} Prior work~\cite{Hanafy:23:War} showed the tension between carbon, energy, and performance in homogeneous clusters. We find that heterogeneity intensifies this conflict. Running a workload on a slower, more energy-efficient server during a high-carbon period might save energy but can ultimately increase carbon emissions, while using a faster, less efficient server on low-carbon periods may lower emissions but waste energy. This matters for heterogeneous environments, such as GPU clusters~\cite{enos2010quantifying, ma2012greengpu}, if carbon efficiency is also a goal alongside energy efficiency. Operators can tune how much they value carbon vs. energy by assigning weights to each.

\noindent\textbf{Limitations.} A key challenge is the complexity of the FJSP. Solvers work for small instances, but complexity grows exponentially with scale. Data center–sized problem settings will need approximation methods or online heuristics with guarantees. Machine learning–based techniques for sustainable systems~\cite{donti2021machine, rolnick2022tackling}, which have accelerated large-scale optimization and stochastic scheduling, may offer a path by learning heuristics that approximate optimal solutions at scale.  
We note that the model we consider in this paper simplifies some real-world conditions. We do not account for several factors that influence carbon reduction, including: Server idle power draw and energy proportionality~\cite{Lefurgy:07, Barroso_warehouse}, limits on task parallelization (i.e., whether a task is truly parallelizable~\cite{fernandez2025hardwarescalingtrendsdiminishing}), and embodied carbon from manufacturing and deployment of servers.

%% file: appendix.tex
\section*{APPENDIX}
\section{Problem Formulation}
\label{sec:problem-formulation}

We model our problem as a variant of the Flexible Job Shop Scheduling Problem (FJSP) with job arrivals, DAG precedence, heterogeneous servers, and time-varying carbon intensity. Our formulation follows the constraint programming style used in prior work~\cite{Park:22}.

The inputs to our problem are as follows:
\begin{itemize}
    \item $\mathcal{J}$: set of jobs; $a_j$: arrival time of job $j$.
    \item $V_j$: set of tasks in job $j$ ($\{t_{j1}, t_{j2},..., t_{jn_j},\}$ ), with $n_j = |V_j|$. 
    \item $G_j = (V_j, \mathcal{E}_j)$: dependency DAG of job $j$ with edges $\mathcal{E}_j$.
    \item $p_{jim}$: processing time of task $i$ of job $j$ ($t_{ji}$) on machine $m$.
    \item $C_{ji}$: completion time of task $t_{ji}$.
    \item $\mathcal{M}$: set of machines and $\mathcal{M}_{ji}$: set of machines that can run task $t_{ji}$.
    \item $P_m$: power draw of machine $m$ (kW).
    \item $I(\tau)$: average carbon intensity at time $\tau$ (gCO$_2$/kWh).
\end{itemize}

A feasible schedule $\mathcal{S}$ assigns each task $t_{ji}$ to a machine and start time, while respecting job arrivals, task dependencies, and machine capacity. For the $i$th task in job $j$, the start time is $s_{j,i}$, and $x_{jim} \in \{0,1\}$ equals 1 if task $t_{ji}$ runs on machine $m$.


The problem is formulated as follows:

\begin{align}
\min \quad & \sum_{t_{ji}}\sum_{m} x_{jim} \sum_{\tau = s_{ji}}^{C_{ji}} P_m \cdot I(\tau), \quad \forall j \in \mathcal{J}, \forall t_{ji} \in V_j, \forall m \in {M_{ji}}\label{obj:carbon}\\
\min \quad & \sum_{t_{ji}}\sum_{m} x_{jim} \, P_m \, p_{jim}, \quad \forall j \in \mathcal{J}, \forall t_{ji} \in V_j, \forall m \in {M_{ji}} \label{obj:energy}\\
\min \quad & \max_{j \in \mathcal{J}} C_{j,n_j} \label{obj:makespan}\\
& s_{ji} \ge a_j, \quad \forall j \in \mathcal{J}, \forall t_{ji} \in V_j \label{eq:arrival}\\
& s_{ji} \ge C_{ji'}, \quad \forall (t_{ji},t_{ji'}) \in \mathcal{E}_j, \forall j \in \mathcal{J} \label{eq:precedence}\\
& \sum_{m \in \mathcal{M}_{ji}} x_{jim} = 1, \quad \forall j \in \mathcal{J}, \forall t_{ji} \in V_j \label{eq:assignment}\\
& C_{ji} = s_{ji} + \sum_{m \in \mathcal{M}_{ji}} x_{jim} \, p_{jim}, \quad \forall (t_{ji},t_{ji'}) \in \mathcal{E}_j, \forall j \in \mathcal{J} \label{eq:completion} \\
& \texttt{NoOverlap} \left(\{[s_{ji}, p_{jim}]: x_{jim}=1\}\right),\label{eq:capacity} \\
& \forall j \in \mathcal{J}, \forall t_{ji} \in V_j, \forall m \in \mathcal{M}_{ji} \notag
\end{align}

The three objectives highlight different optimization goals that the operator can choose from:
(i) \autoref{obj:carbon} minimizes carbon emissions by weighting energy consumption by time-varying carbon intensity, (ii) \autoref{obj:energy} minimizes total energy consumption. (iii) \autoref{obj:makespan} minimizes the makespan (total jobs completion time). The baseline carbon-agnostic scheduler operates based on this objective and only finds the optimal makespan.

When optimizing for carbon or energy objectives, we can impose an upper bound on the makespan. For example, setting $\max_{j \in \mathcal{J}} C_{j,n_j} \le 2 \times C^{\text{OPT}}$ constrains the schedule to finish within twice the optimal makespan $C^{\text{OPT}}$, where $C^{\text{OPT}}$ is obtained from \autoref{obj:makespan}.

\autoref{eq:arrival} ensures tasks cannot start before their job arrives.  
\autoref{eq:precedence} enforces the DAG dependencies between tasks of a job.  
\autoref{eq:assignment} forces each task to be assigned to exactly one server.  
\autoref{eq:completion} defines completion times based on start times and server-dependent processing times.
\autoref{eq:capacity} prevents two tasks from overlapping on the same server, capturing machines' capacity limits.